\documentclass[prl,twocolumn,showpacs,preprintnumbers,amsmath,amssymb]{revtex4}
\newcommand{\ket}[1]{\left | #1 \right\rangle}
\newcommand{\bra}[1]{\left \langle #1 \right |}

\usepackage{amsmath,amssymb,amsthm,times,graphics,graphicx,bm,xcolor}

\begin{document}

\title{Non-monogamy of spatio-temporal correlations and the black hole information loss paradox}

\author{Chiara Marletto and Vlatko Vedral}
\affiliation{Clarendon Laboratory, University of Oxford, Parks Road, Oxford OX1 3PU, United Kingdom and\\Centre for Quantum Technologies, National University of Singapore, 3 Science Drive 2, Singapore 117543 and\\
Department of Physics, National University of Singapore, 2 Science Drive 3, Singapore 117542}
\author{Salvatore Virz\`i}
\affiliation{Universit\`a di Torino, via P. Giuria 1, 10125 Torino, Italy and \\ Istituto Nazionale di Ricerca Metrologica, Strada delle Cacce 91, 10135, Torino, Italy}
\author{Enrico Rebufello}
\affiliation{Politecnico di Torino, Corso Duca degli Abruzzi 24, 10129 Torino, Italy, and\\ Istituto Nazionale di Ricerca Metrologica, Strada delle Cacce 91, 10135, Torino, Italy}
\author{Alessio Avella}
\affiliation{Istituto Nazionale di Ricerca Metrologica, Strada delle Cacce 91, 10135, Torino, Italy}
\author{Fabrizio Piacentini}
\affiliation{Istituto Nazionale di Ricerca Metrologica, Strada delle Cacce 91, 10135, Torino, Italy}
\author{Marco Gramegna}
\affiliation{Istituto Nazionale di Ricerca Metrologica, Strada delle Cacce 91, 10135, Torino, Italy}
\author{Ivo Pietro Degiovanni}
\affiliation{Istituto Nazionale di Ricerca Metrologica, Strada delle Cacce 91, 10135, Torino, Italy}
\author{Marco Genovese}
\affiliation{Istituto Nazionale di Ricerca Metrologica, Strada delle Cacce 91, 10135, Torino, Italy, and\\ INFN, sezione di Torino, via P. Giuria 1, 10125 Torino, Italy}

\begin{abstract}
Pseudo-density matrices are a generalisation of quantum states and do not obey monogamy of quantum correlations. Could this be the solution to the paradox of information loss during the evaporation of a black hole? In this paper we discuss this possibility, providing a theoretical proposal to extend quantum theory with these pseudo-states to describe the statistics arising in black-hole evaporation. We also provide an experimental demonstration of this theoretical proposal, using a simulation in optical regime, that tomographically reproduces the correlations of the pseudo-density matrix describing this physical phenomenon.
\end{abstract}


\maketitle                           



\section{Introduction}
The possibility of black hole evaporation represents a problem from the quantum mechanical perspective \cite{bh,bh1,bh2,bh3}, as well as other cosmological aspects \cite{choud1,mart,malda,choud2}. In short, if the process is unitary as prescribed by quantum theory, then entanglement must be created between the exterior and the interior of the black hole as particle pairs are generated through the process of Hawking radiation \cite{HAW1,HAW2,HAW3,HAW4}. If we provide an elementary model of evaporation based on a finite number of qubits, after half of the qubits in black hole has evaporated, we should presumably have a maximally entangled state between the qubits in the interior and the qubits in the exterior of the black hole, assuming thermal radiation being emitted. As the black hole continues to evaporate, Hawking radiation would imply that even more entanglement is generated between the interior and the exterior of the black hole. But this cannot be, since qubits already maximally entangled cannot be entangled to anything else. This fact, that a system cannot be maximally entangled to more than one other system, is known as the monogamy of entanglement principle \cite{TON}. The claim therefore is that if we are trying to preserve unitarity of black hole evaporation, then the black hole evaporation itself ought to violate monogamy of entanglement \cite{PAGE}.

Here we will not discuss further on the issue of whether there is or is not such a paradox (which has been hotly debated, see e.g. \cite{WAL} and references herein). We would like instead to suggest that, assuming that the paradox exists, a simple re-interpretation of the evaporation process could provide a novel resolution. Informally, this is the rationale for our proposition. Following the Schwarzschild metric, that describes space-time in the presence of a black hole, crossing the horizon for a particle is tantamount to swapping the signatures of the spatial and temporal components of the metric \cite{MIS}. Now, if we think of a typical quantum phase factor $e^{i(kx-\omega t)}$, the change of the sign of space and time simply corresponds to complex conjugation of the phase factor. In this sense, the effect on a density matrix of an in-falling quantum system should be described by the operation of transposition (which swaps the off-diagonal elements and therefore implements the complex conjugation).

It is well known that transposition is a positive, but not completely positive, operation. This means that, if we perform transpose on just one of two entangled systems, the overall state may not end up being a valid density matrix. Here we would like to use this fact to resolve the apparent violation of monogamy of entanglement during the evaporation of a black hole, by suggesting to utilise an extended notion of quantum state to describe it, which includes Hermitean operators that are not positive. These generalised quantum states are called pseudo-density operators (PDOs), \cite{Fit15}.

A density operator can be viewed as a collection of all possible statistics ensuing from measurements of observables of a system of interest. For a $d$-qubit system, for instance, we can write a general density operator as
$$
\rho_d \doteq \frac{1}{2^d} \sum_{i_1=0}^{3}\cdots \sum_{i_d=0}^{3} \langle \bigotimes_{j=1}^{n} \sigma_{i_j} \rangle \bigotimes_{j=1}^{n} \sigma_{i_j}\;.
$$

PDOs generalise these operators into covering statistics that pertain to the time domain.
A general PDO for $d$ qubits is defined as:

$$
R_d \doteq \frac{1}{2^d} \sum_{i_1=0}^{3}\cdots \sum_{i_d=0}^{3} \langle \{ \sigma_{i_j} \}_{j=1}^{n}  \rangle \bigotimes_{j=1}^{n} \sigma_{i_j}\;,
$$

where $\langle \{\sigma_{i_j} \}_{j=1}^{n}  \rangle$ denotes the expected value of a possible set of Pauli measurements which could be either in space or in time, thus generalising standard density operators to cover both space and time correlations.  This is a Hermitian, trace-one but not necessarily positive operator.

Let us understand how the PDO works with an example directly relevant to our problem. Suppose that we want to describe a physical process where a single qubit, initially in a maximally mixed state, is then measured at two different times. Each measurement is performed in all three complementary bases $X, Y, Z$ (represented by the usual Pauli operators). The evolution is trivial between the two measurements, i.e. the identity operator. Suppose now that we would like to write the statistics of the measurement outcomes in the form of an operator, generalising the quantum density operator. Because the whole state, as we said, is Hermitian and unit trace, but not positive, we refer to it as a pseudo-density operator \cite{Fit15}.

It would be represented as:
\begin{equation}
R_{12} = \frac{1}{4} \{I + X_1X_2+Y_1Y_2+Z_1Z_2\}\; ,
\end{equation}
where subscripts 1 and 2 indicate, respectively, the two qubits of a general bipartite state.
This operator looks very much like the density operator describing a singlet state of two qubits, however, the correlations all have a positive sign (whereas for the singlet they are all negative, $\langle XX\rangle = \langle YY\rangle = \langle ZZ\rangle =-1$). In fact, it is simple to show that $R_{12}$ is not a density matrix, because it is not positive (i.e. it has at least one negative eigenvalue).
We can however, trace the label $2$ out and obtain one marginal, i.e. the ``reduced" state of subsystem $1$. Interestingly, this itself is a valid density matrix (corresponding to the maximally mixed state $I/2$). Likewise for subsystem $2$. So, the marginals of this generalised operator are actually both perfectly allowed physical states, but the overall state is not.

Interestingly,

$$
R_{12}=(I\otimes T) \ket{\Sigma}\bra{\Sigma}\;,
$$

where $\ket{\Sigma}=\frac{1}{\sqrt{2}}(\ket{00}+\ket{11})$ and $I\otimes T$ denotes the partial transpose operation. (This relation holds true up to a local bit flip and phase flip for any of the Bell states). Thus, given that the partial transposition can model what happens to a pair entangled qubits due to one of them falling into a black hole, we can use $R_{12}$ as a candidate to describe the state of the pair of qubits, with one of them falling into the black hole.

Based on this heuristic reasoning, we now proceed by proposing a PDO to model the situation where one of the qubits in the pair gets further entangled with a third particle. Specifically, we show that a viable solution to the black hole information problem can be achieved by postulating that a PDO (see eq. \eqref{PDO_BH}), generalising the above pseudo-state $R_{12}$, represents the state of two initially entangled qubits after one of them has crossed the event horizon and fallen into the black hole, getting entangled with a third qubit. As we shall explain, our proposal consists of introducing an extension of the density matrix formalism, via pseudo-density operators, treating equally temporal and spatial correlations. This proposed generalised quantum state can describe perfectly the correlations associated with the black-hole evaporation scenario.

\section{Results}

Suppose that a maximally entangled state is created just above the event horizon of a black hole as in the process of Hawking radiation. One of the particles, e.g. particle 1, now falls into the black hole.  According to our proposal, we conjecture that time-like correlations are created between the two particles (out of what used to be spatial correlations). So the pair is now described by a pseudo-density operator like $R_{12}$, defined above. Now, when another particle, i.e. particle 3, becomes entangled with particle 1, this leads to a three-qubit entangled pseudo-state. In this state, qubits 1 and 2 are maximally temporally correlated, while qubits 1 and 3 are maximally spatially correlated. The total three qubit pseudo-density operator can be written as:
\begin{equation}\label{PDO_BH}
R_{123} = \frac{1}{8} \{I + \Sigma_{12}-\Sigma_{13}-\Sigma_{23}\}\;,
\end{equation}
where $\Sigma_{ij} = X_iX_jI_k+Y_iY_jI_k+Z_iZ_jI_k$. 
The reduced states are $R_{12}= \frac{1}{4}\{I+\Sigma_{12}\}$, $R_{13} = \frac{1}{4}\{I-\Sigma_{13}\}$ and $R_{23} = \frac{I}{4}\{I-\Sigma_{23}\}$.
Now we can see that qubits 1 and 2 can be maximally entangled (in time), while qubits 1 and 3 can also be maximally entangled (in space), as well as qubits 2 and 3. Therefore, correlations described by pseudo-densities need not obey the principle of monogamy of entanglement. We conjecture this pseudo-density operator could be used to describe the elementary step involved in the Black-Hole evaporation.

The usual entanglement monogamy of three qubits can be encapsulated in the following inequality:
\begin{equation}\label{mono}
E_{12} + E_{13} \leq 1 \;.
\end{equation}
This is violated by the state described by $R_{123}$.\\

Therefore, the proposed PDO description for the Black Hole evaporation incorporates the monogamy violation. This is because PDOs, unlike density operators, can be used to describe a situation with two qubits (1 and 2) maximally temporally correlated and one of them forming an entangled state (maximally spatially correlated) with qubit 3. In this framework, the violation of monogamy in Eq. (4) is allowed and predicted. Or course, this requires to modify quantum theory by generalising quantum states from density operators to PDOs. This paper only offers a first exploration of applying this idea to the specific scenario of black-hole evaporation, leaving a more general theory to be developed in the future, should this approach prove useful.

To provide an experimental demonstration of this situation, we performed a quantum optical simulation of such framework. Note that this is not an experimental test, but an illustration of our theoretical proposal within a qubit simulation.
In this experiment, initially, we generate a maximally entangled pair of photons (A and B) in a singlet state.
The correlations between the particle fallen inside the black hole and the one that remained outside, originally belonging to the same maximally entangled state, are observed by measuring photon A at two different times ($t_1$ and $t_2$).
Correlations between the two (spatially) entangled particles inside the black hole, instead, are sensed by measuring photons A and B at the same time $t_1$.
The simulation consists in reconstructing all the relevant statistics contained in the PDO $R_{123}$, by constructing different ensembles of particles.
%
\begin{figure}[ht]
\begin{center}
\includegraphics[width=\columnwidth]{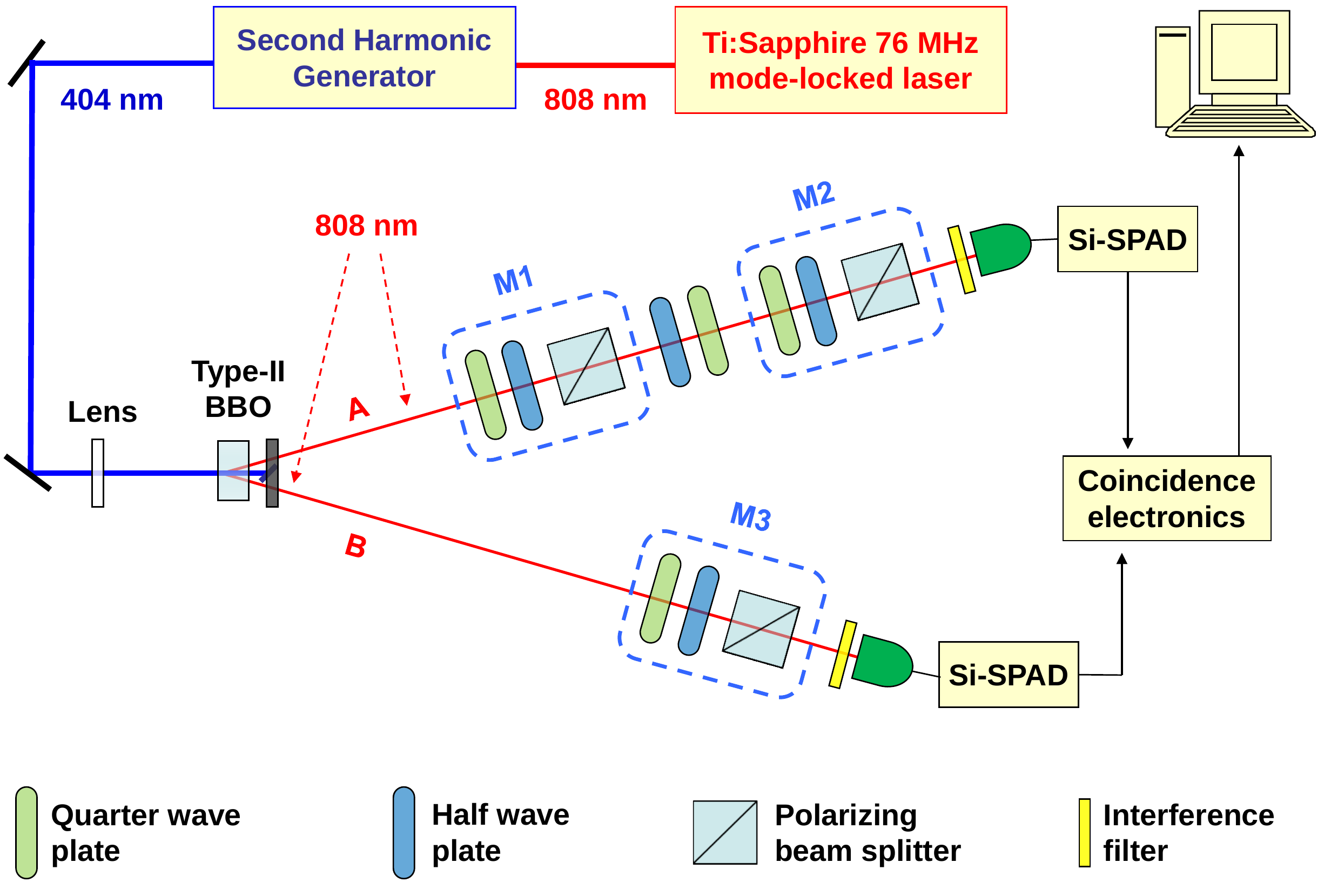}
\caption{Experimental setup. A maximally entangled singlet state is generated by pumping a type-II BBO crystal. Two polarisation measurements, M1 and M2 (at times $t_1$ and $t_2$, respectively) are performed in sequence on photon A, while a single measurement (M3) is carried on photon B. Correlations among them certify entanglement monogamy violation for the whole PDO $R_{123}$ in Eq. (\ref{PDO_BH}), describing the scenario of the spatio-temporal multi-partite entanglement (outside and inside the black hole) considered.}
\label{setup}
\end{center}
\end{figure}
In our setup (see Fig. \ref{setup}), a CW laser at $532$ nm pumps a Ti:Sapphire crystal in an optical cavity, generating a mode-locked pump at 808 nm (repetition rate: 76 MHz) whose second harmonic generation (SHG) is injected into a $0.5$ mm thick $\beta$-barium borate (BBO) crystal to generate type-II parametric down-conversion (PDC) \cite{type-II}.
The maximally entangled singlet state $|\psi_-\rangle=\frac{1}{\sqrt{2}}\left(|HV\rangle-|VH\rangle\right)$ (being $H$ and $V$ the horizontal and vertical polarization components, respectively) is obtained by spatially selecting the photons belonging to the intersections of the two degenerate PDC cones and properly compensating the temporal and phase walk-off \cite{rev}.\\
In photon A path, two polarization measurements occur in cascade (M1 and M2), each carried by a quarter-wave plate (QWP) followed by a half-wave plate (HWP) and a polarizing beam splitter (PBS).
Between the two measurements, a HWP and a QWP are put in order to compensate the polarization projection occurred in M1. Photon B, instead, undergoes a single polarization measurement (M3) performed by the same QWP+HWP+PBS unit used for M1 and M2.
After these measurements, photons A and B are filtered by bandpass interference filters (centered at $\lambda=808$ nm and with a $3$ nm full width at half-maximum) and coupled to multi-mode optical fibers connected to silicon single-photon avalanche diodes (Si-SPADs), whose outputs are sent to coincidence electronics.\\
Initially, we perform an optimized quantum tomographic state reconstruction \cite{tom} on branch A, extracting the temporal correlations allowing to estimate the reduced pseudo-density $R_{12}=\frac{1}{4}(I+\Sigma_{12})$, describing the correlations between particle 1, fallen into the black hole, and particle 2, forming the initial maximally entangled state. To do this, we sum the results obtained in two different acquisitions obtained choosing for M3 orthogonal projectors, e.g. $\ket{H}\bra{H}$ and $\ket{V}\bra{V}$, erasing this way the information on such measurement.
Then, on the spatial side, we measure correlations between M1 and M3 (by having M2 performing the same polarization projection as M1), tomographically reconstructing the reduced pseudo-density $R_{13}= \frac{1}{4}(I-\Sigma_{13})$, corresponding to the generated singlet state $|\psi_-\rangle$, i.e. the one formed by particle 1 with particle 3 within the black hole.\\

\section{Discussion}

\begin{figure}[ht]
\begin{center}
\includegraphics[width=\columnwidth]{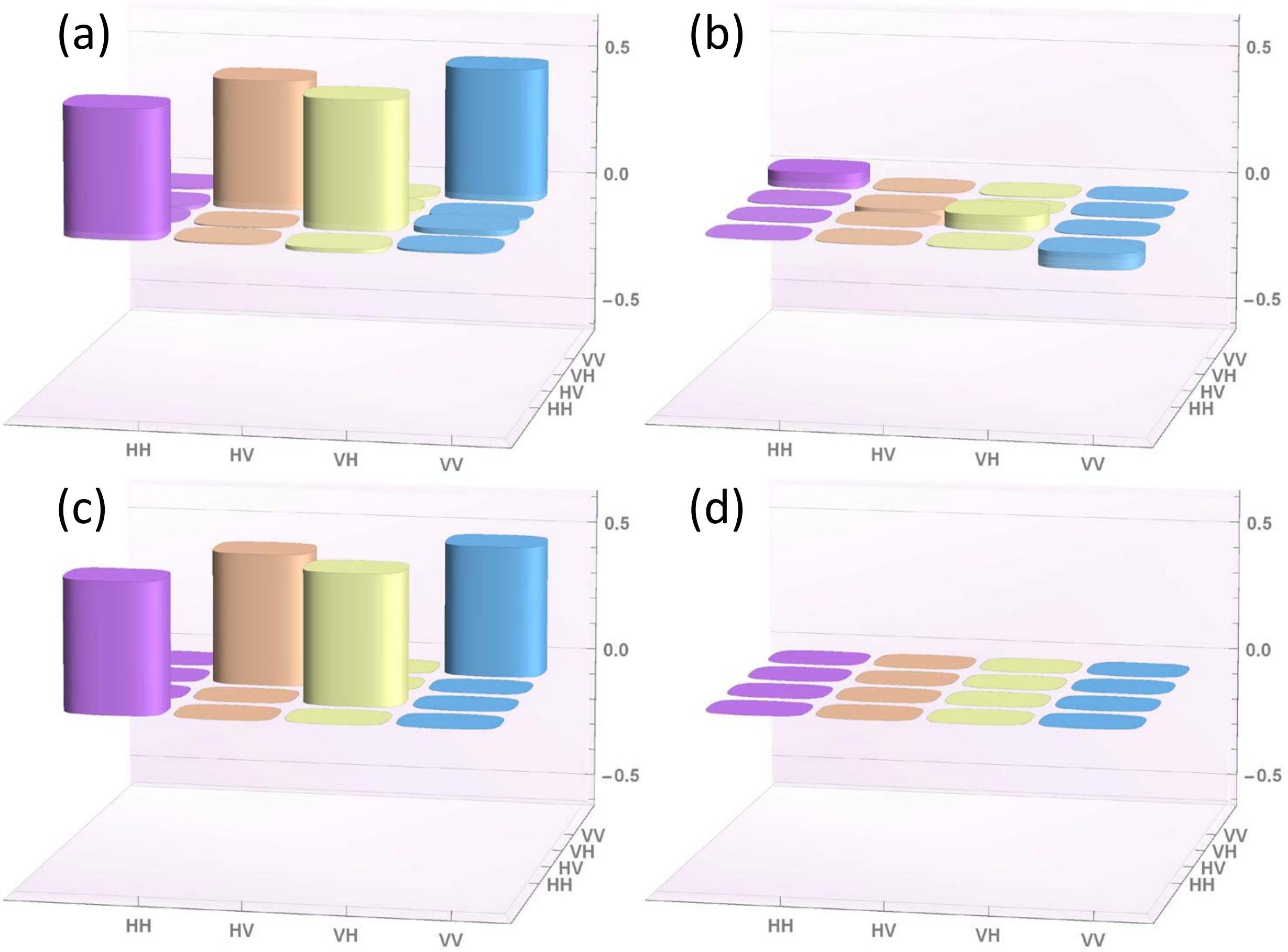}
\caption{Tomographic reconstruction of the real (panel a) and imaginary (panel b) part of the reduced pseudo-density operator $R_{12}=\frac{1}{4}(I+\Sigma_{12})$, describing the temporal correlations between qubits 1 and 2, compared with the corresponding theoretical expectations (panels c and d, respectively).}
\label{R12_BH}
\end{center}
\end{figure}

\begin{figure}[ht]
\begin{center}
\includegraphics[width=\columnwidth]{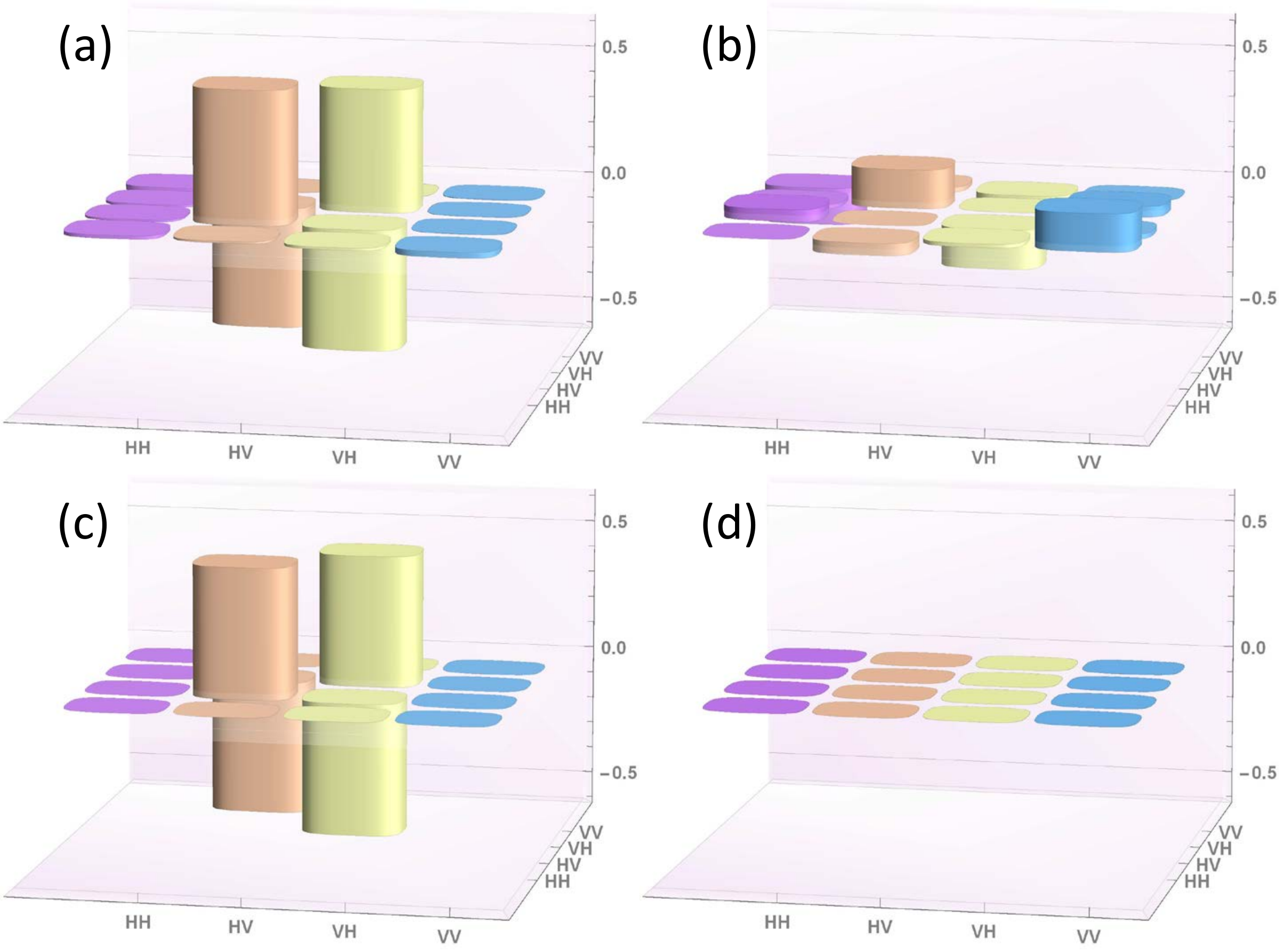}
\caption{Tomographic reconstruction of the real (panel a) and imaginary (panel b) part of the reduced pseudo-density operator $R_{13}=\frac{1}{4}(I-\Sigma_{13})$, related to the spatially maximally entangled state within the black hole, compared with the corresponding theoretically-expected counterparts (panels c and d, respectively).}
\label{R13_BH}
\end{center}
\end{figure}

The results of these two reconstructions are reported in Fig.s \ref{R12_BH} and \ref{R13_BH}, respectively.
In both cases, the experimental results are in excellent agreement with the theoretical expectations, as stated by the Uhlmann's fidelity computed for pseudo-density marginal $R_{13}$ (the only reconstruction corresponding to a physical density matrix), i.e. $F_{13}=96.4\%$.\\
These two reconstructed PDOs would be enough to state the violation of the entanglement monogamy relation reported in Eq. (\ref{mono}), but, as a further test, we experimentally demonstrate the violation of such relation by evaluating the Clauser-Horne-Shimony-Holt (CHSH) inequality \cite{rev} between qubits 1 and 2 (${\rm CHSH}_{12}$) and qubits 1 and 3 (${\rm CHSH}_{13}$).
With the same methodology followed for the quantum tomographic reconstructions of $R_{12}$ and $R_{13}$, we select the proper polarization projections allowing to reach the maximal violation of the CHSH inequality in the temporal domain (M1 and M2) as well as in the spatial one (M1 and M3), obtaining the experimental values ${\rm CHSH}_{12}=2.84\pm0.02$ and ${\rm CHSH}_{13}=2.69\pm0.02$, respectively.
These values grant for the left side of inequality (\ref{mono}) the value:
%
\begin{equation}\label{monoCHSHviol}
E_{12}^{({\rm CHSH})} + E_{13}^{({\rm CHSH})} = 1.380\pm0.009\;,
\end{equation}
being $E_{ij}^{({\rm CHSH})}={\rm CHSH}_{ij}/4$ the amount of entanglement shared between qubits $i$ and $j$,
demonstrating a $42$ standard deviations violation of the entanglement monogamy bound.\\

\section{Conclusions}

In summary, in this paper we propose an alternative resolution of the entanglement paradox in black hole evaporation based on the pseudo-density matrix formalism. We conjectured that the phenomenology of black hole evaporation, as described by Hawking's radiation, could be described by a pseudo-density operator instead of a standard density operator. We propose a specific form for the PDO that could describe a pair of qubits, one falling into a black and the other maximally entangled with a third qubit, after evaporating. The usual paradoxes due to violations of entanglement monogamy do not arise in this formalism, as the PDO can accommodate and describe correlations that violate monogamy. In order to illustrate the temporal and spatial correlations in the proposed PDO, we use a quantum optical demonstration, building an experiment that simulates this physical phenomenon as described by the same pseudo-density matrix. We reconstruct experimentally the correlations in the pseudo-density matrix proposed , demonstrating how violation of entanglement monogamy can emerge in the proposed framework.\\


\section{Acknowledgements}
CM's research was supported by the Templeton World Charity Foundation and by the Eutopia Foundation. VV thanks the Oxford Martin School, the John Templeton Foundation, the EPSRC (UK).
This research is also supported by the National Research Foundation, Prime Minister’s Office, Singapore, under its Competitive Research Programme (CRP Award No. NRF- CRP14-2014-02) and administered by Centre for Quantum Technologies, National University of Singapore.
This research has also been developed in the context of the European Union's Horizon 2020 project ``Pathos''.


\section{References}

\end{document}